\documentclass[aps,prl,twocolumn,groupedaddress]{revtex4}

\usepackage{graphicx}
\usepackage{dcolumn} 
\usepackage{bm}

\begin{document}

\title{Coherent transport of neutral atoms in spin-dependent optical lattice potentials}

\author{Olaf Mandel}
\author{Markus Greiner}
\author{Artur Widera}
\author{Tim Rom}
\author{Theodor W. H\"ansch}
\author{Immanuel Bloch}

\email[Electronic address: ]{imb@mpq.mpg.de} \homepage[URL:
]{http://www.mpq.mpg.de/~haensch/bec}
\affiliation{Ludwig-Maximilians-Universit\"at, Schellingstr.
4/III, 80799 Munich, Germany \\ Max-Planck-Institut f\"ur
Quantenoptik, 85748 Garching, Germany}

\date{\today}
\begin{abstract}
We demonstrate the controlled coherent transport and splitting of
atomic wave packets in spin-dependent optical lattice potentials.
Such experiments open intriguing possibilities for quantum state
engineering of many body states. After first preparing localized
atomic wave functions in an optical lattice through a Mott
insulating phase, we place each atom in a superposition of two
internal spin states. Then state selective optical potentials are
used to split the wave function of a single atom and transport the
corresponding wave packets in two opposite directions. Coherence
between the wave packets of an atom delocalized over up to 7
lattice sites is demonstrated.
\end{abstract}


\maketitle

Over the last years Bose-Einstein condensates in optical lattices
have opened fascinating new experimental possibilities in
condensed matter physics, atomic physics, quantum optics and
quantum information processing. Already now the study of Josephson
junction like effects \cite{Anderson98,Cataliotti01}, the
formation of strongly correlated quantum phases
\cite{Jaksch98,Orzel01,Greiner02a} and the observation of the
collapse and revival of the matter wave field of a BEC
\cite{Greiner02b} have shown some of these diverse applications.
In an optical lattice, neutral atoms can be trapped in the
intensity maxima (or minima) of a standing wave light field due to
the optical dipole force \cite{Grimm00}. So far the optical
potentials used have been mostly independent of the internal
ground state of the atom. However, it has been suggested that by
using spin-dependent periodic potentials one could bring atoms on
different lattice sites into contact and thereby realize
fundamental quantum gates
\cite{Brennen99,Briegel00,Raussendorf01,Brennen02}, create large
scale entanglement \cite{Jaksch99,Briegel01}, excite spin waves
\cite{Sorensen99}, study quantum random walks \cite{Duer02} or
form a universal quantum simulator to simulate fundamental complex
condensed matter physics hamiltonians \cite{Jane02}. Here we
report on the realization of a coherent spin-dependent transport
of neutral atoms in optical lattices
\cite{PhillipsPrivateCommunication}. We show how the wave packet
of an atom that is initially localized to a single lattice site
can be split and delocalized in a controlled and coherent way over
a defined number of lattice sites.

In order to realize a spin dependent transport for neutral atoms
in optical lattices, a standing wave configuration formed by two
counterpropagating laser beams with linear polarization vectors
enclosing an angle $\theta$ has been proposed
\cite{Brennen99,Jaksch99}. Such a standing wave light field can be
decomposed into a superposition of a $\sigma^+$ and $\sigma^-$
polarized standing wave laser field, giving rise to lattice
potentials $V_+(x,\theta)=V_0 \cos^2(kx+\theta/2)$ and
$V_-(x,\theta)=V_0 \cos^2(kx-\theta/2)$. Here $k$ is the wave
vector of the laser light used for the standing wave and $V_0$ is
the potential depth of the lattice. By changing the polarization
angle $\theta$ one can thereby control the separation between the
two potentials $\Delta x =\theta/180^\circ \cdot \lambda_x/2$.
When increasing $\theta$, both potentials shift in opposite
directions and overlap again when $\theta=n\cdot180^\circ$, with
$n$ being an integer. For a spin-dependent transfer two internal
spin states of the atom should be used, where one spin state
dominantly experiences the $V_+(x,\theta)$ potential and the other
spin state mainly experiences the $V_-(x,\theta)$ dipole force
potential. Such a situation can be realized in rubidium by tuning
the wavelength of the optical lattice laser to a value of
$\lambda_x=785$\,nm between the fine structure splitting of the
rubidium D1 and D2 transition. Then the dipole potential
experienced by an atom in e.g. the
$|1\rangle\equiv|F=2,m_F=-2\rangle$ state is given by
$V_1(x,\theta)=V_-(x,\theta)$ and that for an atom in the
$|0\rangle\equiv|F=1,m_F=-1\rangle$ state is given by
$V_0(x,\theta)=3/4 V_+(x,\theta)+1/4 V_-(x,\theta)$. If an atom is
now first placed in a coherent superposition of both internal
states $1/\sqrt{2}(|0\rangle+i|1\rangle)$ and the polarization
angle $\theta$ is continuously increased, the spatial wave packet
of the atom is split with both components moving in opposite
directions. In a fixed polarization configuration, spin-dependent
tunnelling between two lattice sites has been observed in
\cite{Haycock00}.

\begin{figure}
\includegraphics{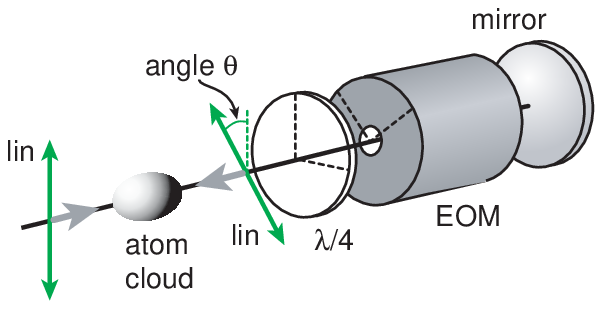}
\caption{\label{fig:expsetup}Schematic experimental setup. A one
dimensional optical standing wave laser field is formed by two
counterpropagating laser beams with linear polarizations. The
polarization angle of the returning laser beam can be adjusted
through an electro-optical modulator. The dashed lines indicate
the principal axes of the wave plate and the EOM.}
\end{figure}

As in our previous experiments, Bose-Einstein condensates of up to
$3\times10^5$ atoms are created in the $|F=1,m_F=-1\rangle$
hyperfine state in a harmonic magnetic trap with almost isotropic
oscillation frequencies of $\omega=2\pi\times 16$\,Hz. A three
dimensional lattice potential is then superimposed on the
Bose-Einstein condensate and the intensity raised in order to
drive the system into a Mott insulating phase \cite{Greiner02a}
and thereby localize the atoms to individual lattice sites. Two of
the three orthogonal standing wave light fields forming the
lattice potential are operated at a wavelength of
$\lambda_{y,z}=840\,$nm. For the third standing wave field along
the horizontal $x$-direction a laser at a wavelength of
$\lambda_x=785\,$nm is used. Along this axis a quarter wave plate
and an electro-optical modulator (EOM) allow us to dynamically
rotate the polarization vector of the retro-reflected laser beam
through an angle $\theta$ by applying an appropriate voltage to
the EOM (see Fig.\,\ref{fig:expsetup}). Initially the polarization
angle $\theta$ is set to a lin$\|$lin polarization configuration.
After reaching the Mott insulating phase we completely turn off
the harmonic magnetic trapping potential but maintain a 1\,G
homogeneous magnetic field along the $x$-direction in order to
preserve the spin polarization of the atoms. This homogeneous
field is actively stabilized to an accuracy of $\approx 1$\,mG.
Shortly before moving the atoms along this standing wave direction
we adiabatically turn off the lattice potentials along the $y$-
and $z$-direction. This is done in order to reduce the interaction
energy, which strongly depends on the confinement of the atoms at
a single lattice site. We can thereby study the transport process
itself, without having to take into account the phase shifts in
the many body state that result from a coherent collisional
interaction between atoms.

By using microwave radiation around 6.8\,GHz we are able to drive
Rabi oscillations between the $|0\rangle$ and the $|1\rangle$
state with resonant Rabi frequencies of $\Omega=2\pi\times
40$\,kHz, such that e.g. a $\pi$-pulse can be achieved in a time
of $12.5\,\mu$s. The microwave field therefore allows us to place
the atom in any of the two internal states $|0\rangle$ or
$|1\rangle$ or an arbitrary superposition of both states.

\begin{figure}
\includegraphics{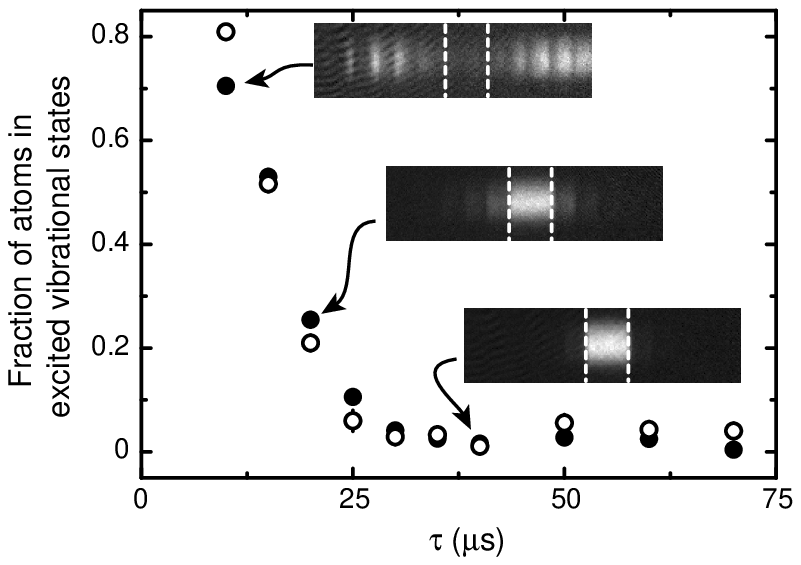}
\caption{\label{fig:excitations} Fraction of atoms in excited
vibrational states after moving the lattice potential in a time
$\tau$ over a distance of $\lambda_x/4$. Filled (hollow) circles
denote atoms in the $|1\rangle$ ($|0\rangle$) state. The images
show the population of the Brillouin zones when the lattice
potential was adiabatically ramped down after the shifting
process. These absorption images correspond to the $|1\rangle$
state and were taken after a time of flight period of 14\,ms. The
white dashed lines in the images denote the borders of the first
Brillouin zone. Atoms within this Brillouin zone correspond to
atoms in the vibrational ground state on each lattice site.}
\end{figure}

During the shifting process of the atoms it is crucial to avoid
unwanted vibrational excitations, especially if the shifting
process would be repeated frequently. We therefore analyze the
timescale for such a movement process in the following way. First
the atom is placed either in state $|0 \rangle$ or state
$|1\rangle$ by using microwave pulses in a standing wave
lin$\|$lin polarization configuration. Then we rotate the
polarization to an angle $\theta=180^\circ$ in a linear ramp
within a time $\tau$, such that again a lin$\|$lin polarization
configuration is achieved. However, during this process the atoms
will have moved by a distance $\pm\lambda_x/4$ depending on their
internal state. In order to determine whether any higher lying
vibrational states have been populated, we adiabatically turn off
the lattice potential within a time of $500\,\mu$s. The population
of the energy bands is then mapped onto the population of the
corresponding Brillouin zones \cite{Kastberg95,Greiner01}. By
counting the number of atoms outside of the first Brillouin zone
of the system relative to the total number of atoms we are able to
determine the fraction of vibrationally excited atoms after the
shifting of the lattice potential (see
Fig.~\ref{fig:excitations}). For a perfectly linear ramp with
infinite acceleration at the beginning and ending of the ramp one
would expect the fraction of atoms in the first vibrational state
to be given by $|c_1(\tau)|^2 = 2 v^2/(a_0 \omega)^2 \sin^2(\omega
\tau/2)$, where $v=\lambda_x/(4 \tau)$ is the shift velocity,
$a_0$ the size of the ground state harmonic oscillator wave
function and $\omega$ the vibrational frequency on each lattice
site.

We have measured the vibrational frequencies on a lattice site for
different polarization angles $\theta$ by slightly modulating the
lattice position and observing a resonant transfer of atoms to the
first excited vibrational state. For atoms in the $|1\rangle$
state the vibrational frequencies remain constant for different
polarization angles $\theta$ as the lattice potential depth
$V_1(x,\theta)$ remains constant. However, for atoms in the
$|0\rangle$ state the lattice potential depth $V_0(x,\theta)$
decreases to 50\% in a lin$\bot$lin configuration. In order to
reduce this effect we tilt the EOM by 3$^\circ$ and thereby
decrease the strength of the $\sigma^-$ standing wave but increase
the strength of the $\sigma^+$ standing wave in such a
polarization configuration. Then both trapping frequencies for the
$|0\rangle$ and the $|1\rangle$ state decrease to approx. 85\% in
a lin$\bot$lin configuration relative to their initial value of
$\omega = 2\pi \times 45$\,kHz in a lin$\|$lin standing wave
configuration. For such trapping frequencies of $\approx$45\,kHz
during the transport process, the excitation probability should
remain below 5\% for shifting times longer than
$\approx2\pi/\omega_x$, taking into account the finite bandwidth
of our high voltage amplifier. This finite bandwidth smoothes the
edges of our linear voltage ramp and thereby efficiently
suppresses the oscillatory structure in the calculated excitation
probability (see Fig.~\ref{fig:excitations}).

\begin{figure}
\includegraphics{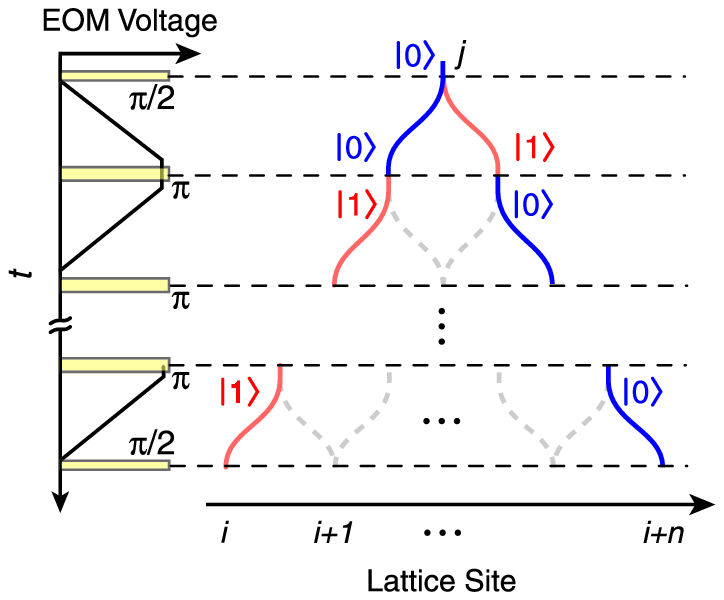}
\caption{\label{fig:interferometer} General interferometer
sequence used to delocalize an atom over an arbitrary number of
lattice sites. Initially an atom is localized to the $j^{th}$
lattice site. The graph on the left indicates the EOM voltage and
the sequence of $\pi/2$ and $\pi$ microwave pulses that are
applied over time (see text).}
\end{figure}

\begin{figure}
\includegraphics{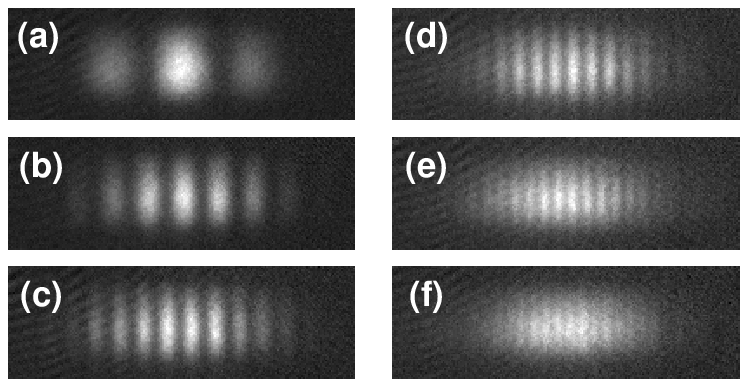}
\caption{\label{fig:interfimages}Observed interference patterns in
state $|1\rangle$ after initially localized atoms have been
delocalized over {\bf (a)} two, {\bf (b)} three, {\bf (c)} four,
{\bf (d)} five, {\bf (e)} six  and {\bf (f)} seven lattice sites
using the interferometer sequence of
Fig.~\ref{fig:interferometer}. The time of flight period before
taking the images was 14\,ms and the horizontal size of each image
is 880\,$\mu$m.}
\end{figure}

\begin{figure}
\includegraphics{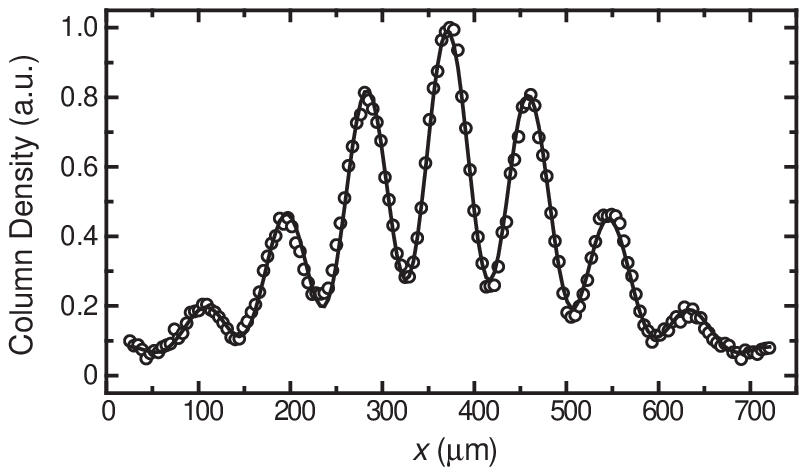}
\caption{\label{fig:interfprofile} Profile of the interference
pattern obtained after delocalizing atoms over three lattice sites
with a $\pi/2$-$\pi$-$\pi/2$ microwave pulse sequence. The solid
line is a fit to the interference pattern with a sinusoidal
modulation, a finite visibility ($\approx$60\%) and a gaussian
envelope. The time of flight period was 15\,ms. }
\end{figure}

In order to verify the coherence of the spin-dependent transport
we use the interferometer sequence of
Fig.~\ref{fig:interferometer}. Let us first consider the case of a
single atom being initially localized to the $j^{th}$ lattice
site. First, the atom is placed in a coherent superposition of the
two internal states $|0\rangle_j$ and $|1\rangle_j$ with a $\pi/2$
microwave pulse (here the index denotes the position in the
lattice). Then the polarization angle $\theta$ is rotated to
180$^\circ$, such that the spatial wave packet of an atom in the
$|0\rangle$ and the $|1\rangle$ state are transported in opposite
directions. The final state after such a movement process is then
given by $1/\sqrt{2} (|0\rangle_j + i\exp(i
\beta)|1\rangle_{j+1})$, where the wave function of an atom has
been delocalized over the $j^{th}$ and the $(j+1)^{th}$ lattice
site. The phase $\beta$ between the two wave packets depends on
the accumulated kinetic and potential energy phases in the
transport process and in general will be nonzero. In order to
reveal the coherence between the two wave packets, we apply a
final $\pi/2$ microwave pulse that erases the which-way
information encoded in the hyperfine states. We then release the
atoms from the confining potential by suddenly turning off the
standing wave optical potential and observe the momentum
distribution of the trapped atoms in the $|1\rangle$ state with
absorption imaging after a time of flight period. As a result of
the above sequence, the spatial wave packet of an atom in the
$|0\rangle$ $(|1\rangle)$ state is delocalized over two lattice
sites resulting in a double slit momentum distribution $w(p)
\propto \exp(-p^2/(\hbar/\sigma_x)^2)\cdot\cos^2(p\,\delta
x_0/2\hbar+\beta/2)$ (see Fig.~\ref{fig:interfimages}a) where
$\delta x_0$ denotes the separation between the two wave packets
and $\sigma_x$ is the spatial extension of the gaussian ground
state wave function on each lattice site. In order to increase the
separation between the two wave packets further, one could
increase the polarization angle $\theta$ to further integer
multiples of 180$^\circ$. In practice, such an approach is however
limited by the finite maximum voltage that can be applied to the
EOM. In order to circumvent this limitation we apply a microwave
$\pi$ pulse after the polarization has been rotated to $\theta =
+180^\circ$, thereby swapping the role of the two hyperfine
states. By then returning the polarization vector to $\theta =
0^\circ$, we do not bring the two wave packets of an atom back to
their original site but rather further increase the separation
between the wave packets (see Fig.~\ref{fig:interferometer}). The
interlaced $\pi$ pulse provides a further advantage of cancelling
inhomogeneous phase shifts acquired in the single particle phase
$\beta$ in a photon echo like sequence. With increasing separation
between the two wave packets the fringe spacing of the
interference pattern further decreases (see
Fig.~\ref{fig:interfimages}). We have been able to observe such
interference patterns for two wave packets delocalized over up to
7 lattice sites (see Fig.~\ref{fig:interfimages}f). When moving
the atoms over up to three lattice sites, the visibility of the
interference pattern remains rather high with up to 60\% (see
Fig.~\ref{fig:interfprofile}). These high contrast interference
patterns directly prove the coherence of the transport process but
also show that the single particle phase $\beta$ acquired for each
atom is almost constant throughout the cloud of atoms in our
system. If the movement process is repeated more often,
inhomogeneously acquired phase shifts over the cloud of atoms
significantly decrease the visibility.

\begin{figure}
\includegraphics{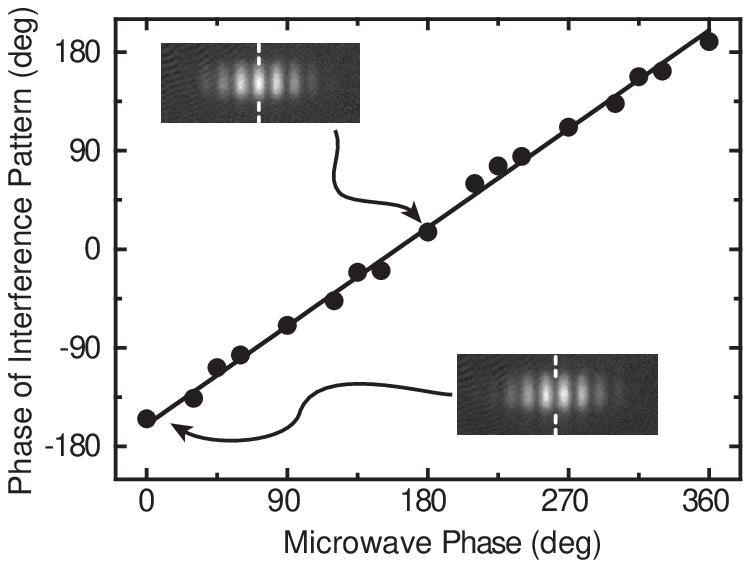}
\caption{\label{fig:interfphase} Phase of the interference pattern
vs the phase $\alpha$ of the final microwave $\pi/2$-pulse in a
$\pi/2-\pi-\pi/2$ delocalization sequence (see
Fig.~\ref{fig:interferometer}). The absorption images show the
measured interference pattern for $\alpha=0^\circ$ and
$\alpha=180^\circ$ after a time of flight period of 15\,ms. The
solid line is a linear fit to the data with unity slope and a
variable offset. The dashed lines in the images correspond to the
center of the envelope of the interference pattern.}
\end{figure}

For many further applications of the coherent spin-dependent
transport it will also be crucial that the single particle phase
$\beta$ is not only constant throughout the cloud of atoms within
a single run of the experiment, but is also reproducible between
different sets of experiments. We have verified this by varying
the phase $\alpha$ of the final microwave $\pi/2$ pulse in a
sequence where an atom is delocalized over three lattice sites. In
Figure~\ref{fig:interfphase} we plot the experimentally measured
phase of the interference pattern vs the phase $\alpha$ of the
final microwave pulse obtained in different runs of the
experiment. We find a high correlation between the detected phase
of the interference pattern vs the phase of the applied microwave
pulse which proves that indeed the single particle phase is
constant between different experiments and can be cancelled via
the phase of the final microwave pulse.

In conclusion we have demonstrated the coherent spin-dependent
transport of neutral atoms in optical lattices, thereby showing an
essential level of coherent control for many future applications.
The method demonstrated here e.g. provides a simple way to
continuously tune the interspecies interactions by controlling the
overlap of the the two ground state wave functions for the two
spin states. Furthermore, if such a transport is carried out in a
three dimensional lattice, where the on-site interaction energy
between atoms is large, one could induce interactions between
almost any two atoms on different lattice sites in a controlled
way. Such controlled interactions of Ising or Heisenberg type
could then be used to simulate the behavior of quantum magnets
\cite{Sorensen99}, to realize quantum gates between different
atoms \cite{Brennen99,Briegel00,Raussendorf01,Brennen02} or to
generate highly entangled cluster states \cite{Jaksch99,Briegel00}
that could form the basis of a one-way quantum computer
\cite{Raussendorf01}.

\begin{acknowledgments}
We would like to thank Ignacio Cirac and Hans Briegel for
stimulating discussions, Alexander Altmeyer for help during the
initial phase of the experiment and Anton Scheich for assistance
with the electronics. We also acknowledge financial support from
the Bayerische Forschungsstiftung.
\end{acknowledgments}

\end{document}